\documentclass{iaus}
\usepackage{graphicx}

\title[Connecting stars to terrestrial planets] 
{Unprecedented accurate abundances: signatures of other Earths?}

\author[J. Mel\'endez, M. Asplund, B. Gustafsson, D. Yong, \& I. Ram\'{i}rez]   
{Jorge Mel\'endez$^1$,
Martin Asplund$^2$,
Bengt Gustafsson$^3$,
David Yong$^4$
 \and Iv\'an Ram\'{i}rez$^2$
 }

\affiliation{$^1$Centro de Astrof\'{i}sica da Universidade do Porto, Rua das Estrelas, 4150-762 Porto, Portugal \\ email: {\tt jorge@astro.up.pt} \\[\affilskip]
$^2$Max-Planck-Institut f\"ur Astrophysik, 
Germany \\[\affilskip]
$^3$Institutionen f\"or fysik och astronomi, Uppsala universitet, 
Sweden \\[\affilskip]
$^4$Research School of Astronomy \& Astrophysics, Australian National University, 
Australia
}

\pubyear{2009}
\volume{265}  
\pagerange{119--126}
\jname{Chemical Abundances in the Universe: Connecting First Stars to Planets}
\editors{K. Cunha, M. Spite \& B. Barbuy, eds.}
\begin{document}

\maketitle

\begin{abstract}
For more than 140 years the chemical composition of
our Sun has been considered typical of solar-type stars.
Our highly differential elemental abundance analysis of 
unprecedented accuracy ($\sim0.01$ dex) of the Sun relative to 
solar twins, shows that the Sun 
has a peculiar chemical composition with a 
$\approx 20$\% depletion of refractory elements relative to the 
volatile elements in comparison with solar twins.
The abundance differences correlate strongly with the 
condensation temperatures of the elements. 
A similar study of solar analogs from planet surveys
shows that this peculiarity also holds in comparisons with solar analogs 
known to have close-in giant planets while the majority of solar analogs
without detected giant planets show the solar abundance pattern. 
The peculiarities in the solar chemical composition can be
explained as signatures of the formation of terrestrial planets
like our own Earth.

\end{abstract}

\firstsection 
\section{Introduction}

For many years people have wondered about how our Sun compares 
to other stars and to whether our existence is related to special 
properties of our solar system.
Angelo Secchi compared the Sun to many bright stars (Secchi 1868). 
He classified the stars in three types: type I which is the modern class A and early F, 
type II which are M stars, and type III comprising modern class G, K and early F, 
and called by Secchi {\it tipo solare} (solar type), due to their 
spectroscopic similarity to the Sun. Furthermore, he concluded that 
``{\it le stelle di questo terzo tipo mostrano di avere una composizione identica 
a quella del nostro Sole}'' (Secchi 1868), meaning that 
solar type stars have identical composition to our Sun. 
Further works (e.g. Payne 1925; Edvardsson et al. 1993; Reddy et al. 2003) 
have not conclusively shown whether the Sun has a normal composition
or not, due to the relatively large ($\gtrsim$ 0.05) remaining systematic
errors (Gustafsson 2008). 
Thus, for more than 140 years father Secchi's conclusion on the 
``universal'' solar composition of the Sun has remained valid.
In order to make further progress, it is important to eliminate many of the
systematic errors ($\sim$ 0.05-0.1 dex) that plague 
stellar chemical composition analyses (Asplund 2005). 
Solar analogs, which are G0-G5 dwarfs, and solar twins, stars
almost identical to the Sun (Cayrel de Strobel 1996), are important in
this context, in particular solar twins, 
because due to their similarity to the Sun a highly differential
analysis will cancel most systematic errors.

Thus, the first step in accurate comparisons of the Sun to other stars
is to find solar twins. After many years of intensive search  
(see review by Cayrel de Strobel 1996), Porto de Mello \& da Silva (1997)
found the first solar twin, 18 Sco [HD 146233], which is much closer 
to the Sun than previous candidates like 16 Cyg B [HD 186427] (see 
e.g. Fig. 2 of Mel\'endez et al. 2006).
More recent surveys have largely increased 
the number of solar twins in the field (Mel\'endez et al. 2006; 
Mel\'endez \& Ram\'{i}rez 2007; Takeda et al. 2007;
Mel\'endez et al. 2009; Ram\'{i}rez et al. 2009) as well as in 
the open cluster M67 (Pasquini et al. 2008).
The most productive survey at high resolution (R $\sim$ 60,000 - 110,000) 
is being undertaken by our group. Most Northern solar twin candidates were 
observed with the 2.7m telescope at the McDonald observatory, 
and complemented with data obtained at the Keck observatory. 
The Southern targets were observed using 
the Magellan Clay 6.5 m telescope at Las Campanas, complemented with
recent (August 2009) VLT observations.

Our solar twin project started in 2002, when only
one solar twin was known.
In order to improve our chances of finding solar twins, we
first expanded the color-temperature relations of Alonso et al. (1996)
to other photometric systems (Mel\'endez \& Ram\'{i}rez 2003), 
allowing us to use existing photometry in other systems
(e.g. Geneva) to select the best targets. We later improved
the calibrations adding more stars and including new 
homogeneous systems like Tycho ($B_T - V_T$) and 2MASS 
(Ram\'{i}rez \& Mel\'endez 2005).
Although our first solar twin proposal in 2004 did not fly, 
the same year a more ``exciting'' proposal
on Li in halo stars
was granted a few nights with Keck.
A small amount of that observing time was
devoted to twin candidates, resulting in the
discovery of the second solar twin (HD 98618; Mel\'endez et al. 2006) 
and revealing that our temperature scale
(and that of Alonso et al. 1996),
although precise, have probably a zero-point issue. 
Our new selection of candidates from the Hipparcos catalog
took into account a preliminary zero-point offset, 
which has been recently confirmed by analyses of
solar twins and used for improved temperature calibrations 
(Casagrande et al. 2009).

New solar twin proposals at McDonald and Magellan at Las Campanas 
(through Australian access) during 2006 were successful.
The first observing run at McDonald in April 2007 allowed us
to identify the best solar twin known to date, HIP 56948 (Mel\'endez \& Ram\'{i}rez 2007),
which is not only very similar to the Sun physically, 
but has also a low Li abundance similar to solar.
Its status of best solar twin has been recently confirmed by Takeda \&  Tajitsu (2009).
On the other hand, the Magellan observations at Las Campanas 
are opening new windows for astrophysics
of the 0.01 dex level in chemical abundance accuracy.
New observations taken recently at the VLT with UVES and CRIRES, 
promise to achieve even better precision (0.005 dex, $\sim$1\%),
and to use solar twins to set tight constraints on Li and Be 
depletion in the Sun (e.g. do Nascimento et al. 2009).

\begin{figure}
\begin{center}
\includegraphics[width=3.4in]{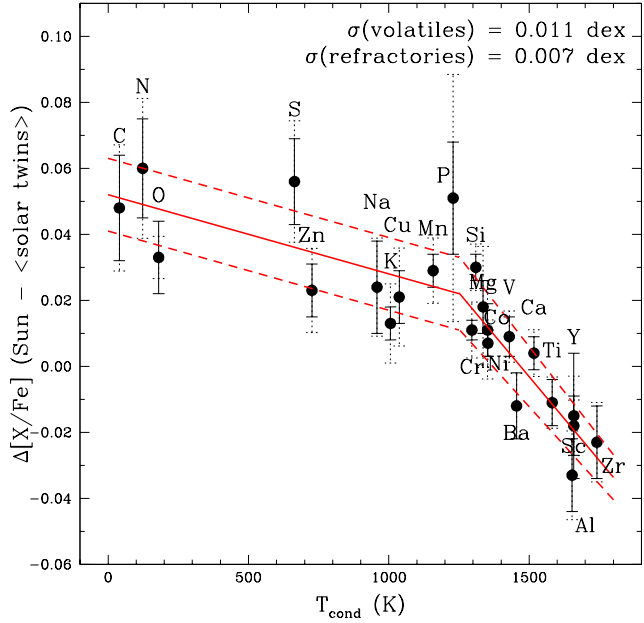} 
 \caption{Differences between [X/Fe] of the Sun and the 
mean values in the solar twins 
as a function of $T_{\rm cond}$. 
The abundance pattern shows a break at $T_{\rm cond}$ $\sim 1200-1250$ K. 
The solid lines are fits to the abundance pattern, while the dashed lines represent the 
standard deviation from the fits. 
The low element-to-element scatter from the 
fits for the refractory 
($\sigma = 0.007$ dex) and volatile ($\sigma = 0.011$ dex) elements 
confirms the high precision of our work.
Observational errors (including errors in both the Sun and solar twins) 
are shown with dotted error bars, while the errors in the mean 
abundance of the solar twins are shown with solid error bars.
}
   \label{fig1}
\end{center}
\end{figure}

\section{Terrestrial planet signatures}

Our observations taken with the MIKE spectrograph and
the 6.5m Magellan Clay telescope at Las Campanas, 
have definitely shown, for the first time, that the Sun has
a peculiar chemical composition (Mel\'endez et al. 2009).
As can be seen in Fig. 1, the difference in chemical abundances
between the Sun and the mean value of the solar twins does not
cluster around zero. Instead, the abundance differences have a
remarkable correlation with the condensation temperature of the
elements (Lodders 2003). This peculiarity has been recently verified 
by Ram\'{i}rez et al. (2009) using McDonald observations.

A fascinating possibility to explain the chemical anomalies of the Sun 
is that they are related to its properties as a host of terrestrial planets. 
A particularly striking circumstance is that the inner solar system planets 
and meteorites are enriched in refractories compared to volatiles (e.g. Alexander
et al. 2001), with a break at $T_{\rm cond}$ $\sim$1200 K, identical to the
break at $T_{\rm cond}$ $\sim$1200-1250 K detected in the solar abundance pattern (Fig. 1). 
The abundance pattern of meteorites is a mirror-image of the solar 
pattern relative to the solar twins.
Also, a radial gradient exists, with greater enhancement of refractories at 
smaller heliocentric distances (Palme 2000).
The break in the chemical abundance pattern at $T_{\rm cond} \approx 1200$ K (Fig . 1), 
suggests that the volatiles retained their original abundances both in the Sun 
and the solar twins. Such temperatures are only encountered in the inner parts ($<3$ AU) 
of proto-planetary disks (Ciesla 2008), which also suggests that the abundance 
pattern is related to the presence of terrestrial planets. 
Furthermore, iron meteorites (like those we see in museums around the world)
were probably formed in the terrestrial planet region and 
may be a remnant  of Earth forming material (Bottke et al. 2006),
hence being perhaps formed from the iron missing in the Sun.

The amount of dust-depleted gas required to explain the solar abundances 
is similar to the mass of refractories locked up in 
Mercury, Venus, Earth and Mars (Mel\'endez et al. 2009).
Thus, it is tempting to speculate that the formation of the terrestrial 
planets might have given the Sun its special surface composition. 
The disk masses during the T Tauri phase are $\sim 0.02$ $M_{\odot}$ but 
larger values are possible, providing thus enough material 
to change the solar photospheric composition. 
However it may be a problem with time-scales since
proto-planetary disks are observed to have typical 
life-times $<10$ Myr (Sicilia-Aguilar et al. 2006), 
but the deep solar convection zone
at that time would have erased the planetary signatures,
unless the proto-solar nebula was abnormally long lived 
so that the dust-depleted gas was accreted $\sim$20 Myr later when the
solar convection zone was thin. 
Another possibility is that the early Sun was never fully convective,
as shown by the dynamical star-formation 
calculations with a time-dependent convection treatment
of \cite{wuchterl03} and 
\cite{wuchterl01}, making 
it easier to imprint the signatures of other Earths (Nordlund 2009).

We have also analyzed a sample of 
10 solar analogs from planet surveys, four having inner giant planets, 
while for the other six no planets have been detected yet. 
Our analysis shows that all solar analogs with giant planets differ from the Sun 
but closely resemble most solar twins. 
Thus, the odd solar composition is not due to giant planets as such,
but probably related to terrestrial planets. 
Indeed, only two of the six solar analogs without close-in giant planets have 
abundances that differ significantly from the solar pattern. 
The fraction of stars with the solar pattern seems thus
tentatively related to the presence of giant planets on close orbits: $\approx 0$\% 
when having such planets and $\approx 10-20$\% for solar-type stars in general,
and $\sim$50\%-70\% without close-in giant planets.
Although the statistics has to be improved considerably, 
the numbers are clearly tantalizing. 

\section{How typical is our solar system?}
Although the samples are still relatively small, considering
both the Magellan (Mel\'endez et al. 2009) and McDonald (Ram\'{i}rez
et al. 2009) samples, we conclude that
about 10-20\% of stars \emph{physically} 
similar to the Sun have also \emph{chemical} similarities to
our Sun and therefore they
may be hosting terrestrial planets like our own Earth.

A complete high precision spectroscopic survey of solar twins 
is urgently needed in order to know which stars may be 
the best candidates of hosting solar system twins and 
to improve the statistics on the fraction of stars 
hosting other Earths and perhaps life.

\end{document}